\documentclass[12pt]{iopart}
\usepackage{graphicx}
\usepackage{epstopdf,ulem}
\usepackage{color}
\usepackage{xcolor}
\usepackage{epsfig,esint}
\usepackage{amsfonts}
\usepackage{amssymb}
\usepackage[utf8]{inputenc}
\usepackage{braket}
\usepackage{bm}
\usepackage{dcolumn}                 

\def\thesection{\arabic{section}}
\def\thetable{\arabic{table}}

\def\beq{\begin{equation}}
\def\eeq{\end{equation}}
\def\beqa{\begin{eqnarray}}
\def\eeqa{\end{eqnarray}} 
\def\ban{\begin{eqnarray*}}
\def\ean{\end{eqnarray*}}
\def\bi{\begin{itemize}}
\def\ei{\end{itemize}}

\def\f{\frac}

\def\beq{\begin{equation}}
\def\eeq{\end{equation}}
\def\vep{\tilde {\varepsilon}}
\def\ep{\epsilon}

\def\f{\frac}
\def\bs{\boldsymbol}

\usepackage{subfiles} 

\begin{document}

\title[Pairing effects in nuclear pasta phase...]
{Pairing effects in nuclear pasta phase within the relativistic Thomas-Fermi formalism}
\author{U J Furtado$^{1}$, S S Avancini$^{1}$ and J R Marinelli$^{1}$}
\address{$^1$ Depto de F\'{\i}sica - CFM - Universidade Federal de Santa
	Catarina  Florian\'opolis - SC - CP. 476 - CEP 88.040 - 900 - Brazil.}
\ead{ujfurtado@gmail.com}

\begin{abstract}
Pairing effects in non-uniform nuclear
matter, surrounded by electrons, are studied in the
protoneutron star early stage and in other conditions.
The so-called nuclear
pasta phases at subsaturation densities are solved in a Wigner-Seitz cell, within
the Thomas-Fermi approximation. The solution of this problem is important for the
understanding of the physics 
of a newly born neutron star after a supernova explosion. It is shown
that the pasta phase is more stable than uniform nuclear matter on some conditions and
the pairing force relevance is studied in the determination of these stable phases.
\end{abstract}
\noindent{\it Keywords\/}: {Nuclear pairing, pasta phase, Thomas-Fermi, Wigner-Seitz, Supernovae, neutron star.

\vspace{0.50cm}

This Accepted Manuscript is available for reuse under a CC BY-NC-ND licence after the 12 month embargo period provided that all the terms and conditions of the licence are adhered to.}


\section{Introduction}\label{intro}

It is well established that exotic inhomogeneous structures named nuclear pasta phase are more stable than uniform nuclear matter for matter in supernovae and neutron stars at subsaturation densities ($0.1\rho_0 \lesssim \rho \lesssim 0.8 \rho_0$), where $\rho_0$ is the (symmetric) nuclear saturation density ($\rho_0 \sim$ $2.3\times 10^{14}$ g/cm$^3$ $\sim$ 0.16 fm$^{-3}$)
\cite{Ravenhall1983b,Hashimoto1984,Caplan2018}, and it is expected to exist until temperatures of $k_B T\approx 15$ MeV before melting \cite{Avancini2009,Schuetrumpf2013}.
The pasta phase has been often calculated in the literature for electrically neutral Wigner-Seitz (WS) cells of appropriate geometries, containing 
neutrons, protons and electrons. Both relativistic and non-relativistic mean-field 
calculations have been done.
Thomas-Fermi (TF) approximation \cite{Avancini2009,Maruyama2005} has been used,
but Hartree-Fock calculations \cite{Magierski2002,Grygorov2010}
have been used as well, all within the  Wigner-Seitz approximation.
Other approaches to this problem, that go beyond the WS approximation, are based on classical \cite{Horowitz69-2004,Horowitz2005,Horowitz2015,Schneider2013} or quantum molecular dynamics \cite{Watanabe2005,Sonoda2008,Watanabe2009}.
The pasta has been also treated within the Coexisting Phase Approximation \cite{Avancini2008,Barros2020} and parameterized density profiles for Skyrme functionals are also found \cite{Pearson2020}.

The supernova explosion mechanism has been a long-standing problem in astrophysics and is an issue that has been attracting the attention of many research groups. Nonetheless, it is still at present not  a completely solved problem due to its tremendous complexity \cite{Janka:2016fox,Rezzolla:2018jee}. 
However, the fundamental role of the neutrino transport properties seems to be uncontroversial for the understanding of supernova explosions. 
After the death of a massive star, the core-collapse supernova (CCSN) initiates and follows several stages, being the detailed description model dependent, and ultimately ends producing a neutron star (NS) or a black hole.

Roughly speaking, the core collapses under gravity and when it reaches the density of nuclear matter, the equation of state stiffens, the collapse halts, the matter in the inner region bounces, and a protoneutron star (PNS) is formed.  The bouncing inner core drives a shock wave  propagating in the direction of the supersonic infalling outer core region which weakens and stalls. Subsequent heating by neutrino diffusion may cause 
a shock revival and an explosion. After the successful CCSN explosion, the PNS cooling and deleptonization is determined by neutrino emission  until finally a neutron star is born.

The key point is that the ``pasta phase''
is expected to be formed  in the low-density regions of both the NS inner crust  and the supernova environment. This inhomogeneous matter will have a prominent role in the transport and elastic properties of the neutron star and supernova matter, hence, being crucial in determining their properties \cite{Sonoda:2007ni}.
The existence of this inhomogeneous matter  strongly affects the neutrino transport properties in the CCSN \cite{Horowitz70-2004,Furtado2016}, which may have an important role in the neutrino opacities, which in turn is an important issue for the description of 
the dynamics of the core collapse and the posterior cooling and deleptonization
of the proto-neutron star.
Thus, the complete study of the pasta phase properties is necessary for a 
definitive understanding of the  CCSN and the subsequent neutron star formation. 
In the inner crust of a neutron  star, the pasta phase is
expected to be  formed in a density range of the order of $0.1\rho_0 \lesssim \rho \lesssim 0.8\rho_0$, the matter is essentially cold and the proton fraction is small ($y_p \sim 0.1$). The X-ray emission, its cooling, pulsar glitches, gravitational waves, oscillation modes, are properties that depend on the structure of the NS crust.

In the supernova environment until the final phase of the supernova collapse
just before the core bounce, the pasta phase is believed to be present at subsaturation densities with larger proton fractions ($y_p \sim 0.3$) compared to the neutron star matter \cite{Watanabe:2011}. 
Also, when the cooling of the PNS proceeds
the pasta formation is possible. 

One can see in \cite{Pons1999,Hempel2012} that conditions of temperature, nuclear density, and proton fraction $y_p$ consistent with the existence of nuclear pasta may exist in supernovae from a few milliseconds before the bounce, where $y_p\approx 0.3$, and after that, mainly in the outer layers, with the proton fraction decreasing to about $y_p\approx 0.1$.

The pasta is expected to 
have an important role in the properties of strongly magnetized neutron stars, named magnetars \cite{Duncan:1992hi}. Their relatively large spin period has been attributed \cite{Pons:2013nea} to effects due to the presence of the pasta phase in the inner crust of such stars \cite{Fang2017}.

The dynamics in the neutron star crust may be profoundly 
altered, as happens in terrestrial experiments, due to the superfluidity phenomena. The shear and bulk viscosity are strongly affected by the presence of superfluid neutrons \cite{Manuel2011}.
The specific heat of the pasta phase was shown to be very sensitive to pairing properties \cite{Barranco1998,Pastore2014a,Pastore2014b}, significantly affecting the thermalization of neutron star crusts \cite{Fortin2010}. Besides the specific heat, thermalization is also influenced by neutrino emissivity, and pairing effects are expected to be important in this subject too \cite{Burrello2016,Sedrakian:2018ydt}.

An important factor in NS dynamics is the appearance of an array of quantized vortices.
The interaction between them and other components of the star may cause a dissipative
phenomena, which may explain sudden spin-up periods (or ``glitches'') that are observed
in some pulsars, like in Vela star \cite{Anderson1975,Pines1985}. On the other hand, it is a well known fact
that there is a class of NS that are highly magnetized objects and the corresponding
magnetic field is modified by the presence of superconductivity and vice-versa. According
to \cite{Sinha2015}, the size
of the magnetic field in the core of the NS can affect the superconducting region. The
consequences of the nucleonic pairing on the phenomenology of magnetars were discussed
in \cite{Sedrakian2017}. Another possible effect is related to the
rapid cooling observed in Cassiopeia A star, which is believed to occur due to the opening of
a new channel for neutrino emission as a consequence of neutron superfluidity \cite{Page2011},
although other possible interpretations can be done to this cooling \cite{Blaschke2013,Sedrakian2013}. We didn’t go here to the details of the relationship between
pairing and those possible observational phenomena, most of which can be seen in \cite{Haskell2018}.

In the literature, the pairing has been included in nuclear matter via Hartree-Fock-Bogolyubov (HFB) \cite{Decharge1980}, via a BCS gap equation in a relativistic mean-field theory (RMF) for some pairing interaction \cite{Gambhir1990,Ring1996}, and also via relativistic HFB \cite{Kucharek1991}.

Although in supernovae the temperature is finite, of about a few MeV's in the region of interest, we considered here zero temperature in a first investigation, where the effect of the pairing is supposed to be maximum \cite{Goodman1981}. The proton fractions we considered here are $y_p=0.3$ and $y_p=0.1$.

It was shown in \cite{Okamoto2012} and in the works based on molecular dynamics that, by performing pasta calculations taking large enough cells to include several units of pasta structures, different structures from the usual ones or mixed states may appear close to the transition densities between different pasta structures. Despite that, we considered here only one kind of structure (droplet, rod, slab, tube or bubble) for each density, the one that provided the lower energy.

The present paper deals essentially with the study of the inner crust of a NS, which is believed to consist of a neutron-rich matter formed by clusters of nucleons embedded in a gas of free electrons and dripped neutrons arranged in a crystalline lattice. Here, we followed the approach often used in the literature, where the infinite periodic crystalline lattice is treated in the Wigner-Seitz approximation, i.e.,  by one independent, non-interacting, and electrically neutral cell (WS unit cell) containing the nuclear cluster in its center. The Wigner-Seitz approximation has been shown to be a very good approximation in most of the inner crust region, except in the deeper region of the crust close to the crust-core transition \cite{Newton2009}.
We considered WS cells of different geometries in order to minimize the energy of the cell in the relativistic Thomas-Fermi approximation. The Coulomb lattice energy was included self-consistently in the calculation.
As long as the density increases WS-cells of spherical-like, rod-like, slab-like, tube-like, and bubble-like geometries (pasta phases) are considered in order to look for the minimum energy of the system.

Here we followed the Thomas-Fermi approach as described in \cite{Avancini2008}
in order to generate the pasta phase structures, starting from a field
theoretical approach. The pairing was included self-consistently within this approach, following the procedure given in \cite{Tian2009}. The self-consistent calculation was
performed considering matter with fixed proton fraction, for the values mentioned above, where only protons, neutrons and electrons were present. Besides the equation of state properties we also show some results for the specific heat of the pasta phase \cite{Barranco1998,Pastore2014a,Pastore2014b,Margueron2012}.

To conclude this section, we quote some recent calculations of the NS inner crust using the TF approximation that are complementary to ours. In \cite{Pearson2020} the extended Thomas-Fermi (ETF) calculation using a Skyrme-like functional without the pairing correction is performed. A similar analysis  including the pairing interaction also in an ETF calculation and using several non-relativistic Skyrme-like functionals is performed in \cite{Shelley2021}. 
In \cite{Ji2021} the nuclear pasta is studied using the TF formalism in a fully three-dimensional calculation using a relativistic mean-field approach with the point-coupling interaction.  In \cite{Shen2020}  the calculation of the NS inner crust is performed using the relativistic TF approximation without the pairing interaction. This latter calculation assumes a parameterized form for the neutron and proton densities and a uniform background electron gas in the neutral WZ cell. Finally, in \cite{Sharma2015} a self-consistent TF calculation based on a microscopic energy density functional is used for the calculation of the inner crust.  In this last calculation, the pairing interaction is not added for the calculation of the inner crust.

\section{Formalism}\label{formalismo}

In this Section, the energy of a Wigner-Seitz (WS) cell within the Thomas-Fermi (TF) approximation, based on a Lagrangian density used to describe the nuclear matter within the cell, is briefly presented, as well as the formalism used to include the pairing interaction between the nucleons.

\subsection{Thomas-Fermi energy}\label{ssec energia}

The Lagrangian density for a system composed of protons, neutrons and electrons based in the well-known non-linear
Walecka model is \cite{Avancini2008,Walecka1986,Glendenning2000}:
\begin{equation}
\mathcal{L}=\sum_{i=p,n}\mathcal{L}_{i}\mathcal{\,+L}_{{\sigma }}\mathcal{+L}_{{\omega }}
\mathcal{+L}_{{\rho }}\mathcal{+L}_{{\omega \rho }}
\mathcal{+L}_{{\gamma }}\mathcal{\,+L}_{{e }},
\label{eq lag}
\end{equation}
where the nucleon and electron Lagrangian densities are:
\begin{eqnarray}
&\mathcal{L}_{i}=\bar{\psi}_{i}\left[ \gamma _{\mu }iD^{\mu }-M^{*}\right]
\psi _{i}  \label{eq lagnucl}; \\
&\mathcal{L}_{e}=\bar{\psi}_{e}\left[ \gamma _{\mu }(i\partial ^{\mu }+eA^{\mu})-m_e\right]
\psi _{e},
\end{eqnarray}
with
\begin{eqnarray}
&iD^{\mu } =i\partial ^{\mu }-g_vV^{\mu }-\frac{g_\rho}{2}{\bs{\tau}}
\cdot \mathbf{b}^{\mu } - e \frac{1+\tau_3}{2}A^{\mu}; \label{eq Dmu} \\
&M^{*} =M-g_s\phi.
\label{eq Mstar}
\end{eqnarray}

The meson (sigma, omega and rho, respectively) and electromagnetic Lagrangian densities are
\begin{eqnarray}
&\mathcal{L}_{{\sigma }} =\frac{1}{2}\left( \partial _{\mu }\phi \partial %
^{\mu }\phi -m_{s}^{2}\phi ^{2}-\frac{\kappa}{3} \phi^3 -\frac{\lambda}{12} \phi^4 \right);  \\
&\mathcal{L}_{{\omega }} =\frac{1}{2} \left(-\frac{1}{2} \Omega _{\mu \nu }
\Omega ^{\mu \nu }+ m_{v}^{2}V_{\mu }V^{\mu }+ \frac{\zeta}{12}g_v^4 (V_\mu V^\mu)^2 \right); \\
&\mathcal{L}_{{\rho }} =\frac{1}{2} \left(-\frac{1}{2}
\mathbf{B}_{\mu \nu }\cdot \mathbf{B}^{\mu
\nu }+ m_{\rho }^{2}\mathbf{b}_{\mu }\cdot \mathbf{b}^{\mu } \right);\\
&\mathcal{L}_{{\gamma }} =-\frac{1}{4}F _{\mu \nu }F^{\mu \nu };\\
&\mathcal{L}_{{\omega \rho }} =\Lambda (g_\rho^2 \mathbf{b}_{\mu }\cdot \mathbf{b}^{\mu})(g_v^2(V_\mu V^\mu)) ,
\end{eqnarray}
where $\Omega _{\mu \nu }=\partial _{\mu }V_{\nu }-\partial _{\nu }V_{\mu }$
, $\mathbf{B}_{\mu \nu }=\partial _{\mu }\mathbf{b}_{\nu }-\partial _{\nu }\mathbf{b}%
_{\mu }-g_\rho(\mathbf{b}_{\mu }\times \mathbf{b}_{\nu })$ and $F_{\mu \nu }=\partial 
_{\mu }A_{\nu }-\partial _{\nu }A_{\mu }$. The electromagnetic coupling constant is given by $e=\sqrt{4 \pi/137}$
 and $\bs{\tau}$ is the isospin operator.
We also included a mixed isoscalar-isovector coupling $\Lambda$ to allow models parametrizations like FSUGold \cite{Todd-Rutel2005} to be considered.
The parametrizations we used are NL3 \cite{lalazissis1997}, FSUGold, FSU2H and FSU2R \cite{Tolos2017}.
All the parametrizations used in our calculations support a two-solar masses neutron star, except for FSUGold. A detailed description of the properties of these models is given in \cite{Dutra:2014qga,Tolos:2016hhl}.
 
As a first investigation, we considered zero temperature. The Euler-Lagrange equations were solved in the self-consistent mean field approach within the TF approximation \cite{Avancini2008,Avancini:2010ch}.
In the TF approximation the field operator mean-values for nucleons and electrons are
identified with their respective densities as follows \cite{Avancini2008}:
\begin{eqnarray}
&\rho(\bs{r})= \rho_p(\bs{r}) + \rho_n(\bs{r})= \left\langle
\hat{\psi}^{\dagger} \hat{\psi} \right\rangle; \\
&\rho_s(\bs{r})= \rho_{sp}(\bs{r}) + \rho_{sn}(\bs{r})= 
\left\langle \hat{\bar{\psi}} \hat{\psi} \right\rangle; \\
&\rho_3(\bs{r})= \rho_p(\bs{r}) -\rho_n(\bs{r}) = 
\left\langle  \hat{\psi}^{\dagger} \tau_3 \hat{\psi} \right\rangle; \\
&\rho_e(\bs{r})= \left\langle  \hat{\psi}^{\dagger}_e  \hat{\psi}_e \right\rangle,
\end{eqnarray}
where
\begin{eqnarray}
&\rho_i(\bs{r}) = \f{\gamma}{(2\pi)^3}\int_0^{\bs{k}_{Fi}(\bs{r})}  d^3k ,\quad i=p,n,e; \label{eq eq dens b} \\
&\rho_{si}(\bs{r}) = \f{\gamma}{(2\pi)^3}\int_0^{\bs{k}_{Fi}(\bs{r})} \f{M^*}{E^*}~d^3k, \quad i=p,n, \label{eq dens s}
\end{eqnarray}
are the particle and scalar densities respectively, which are position dependent,
$\gamma=2$ is the spin multiplicity and $\bs{k}_{Fi}$ is the Fermi momentum. The fields and the densities are position dependent now, but for aesthetic reasons we will not always explicitly write this dependence.

With the approximations, the energy density is:
\begin{eqnarray}
\Sigma=&\sum_{i=p,n,e} \ep_i
+ g_v V_0 \rho_B + \f{1}{2} g_\rho b_0 \rho_3 + e A_0 (\rho_p-\rho_e) \nonumber \\
&- \left[- \f{1}{2}(\nabla \phi_0)^2-\f{1}{2}m_s^2 \phi_0^2-\f{1}{2} \left (\f{\kappa}{3} \phi_0^3 + \f{\lambda}{12} \phi_0^4 \right) \right. \nonumber \\
&+ \f{1}{2}(\nabla V_0)^2 + \f{1}{2} m_v^2 V_0^2 + \f{\zeta}{24}g_v^4 V_0^4 + \Lambda g_\rho^2 \, b_0^2 \, g_v^2 V_0^2\nonumber \\
&\left. + \f{1}{2}(\nabla b_0)^2 + \f{1}{2}m_\rho^2 b_0^2
+ \f{1}{2}(\nabla A_0)^2 \right],
\end{eqnarray}
with
$$
\ep_i(\bs{r}) = \f{\gamma}{(2\pi)^3}\int_0^{\bs{k}_{Fi}(\bs{r})} E_i^* ~d^3k,
$$
$E_i^*=\sqrt{k^2+m_i^2}$ and $m_i=M^*$ for the hadronic particles and $m_i=m_e$ for the electron. In the TF approach the system is considered locally uniform, which justify the above expressions
for the particle kinetic energies and densities. The space integration of $\Sigma$ over a cell gives the total energy of the cell, $E_{TF}$.

The field equations can be written as:
\begin{eqnarray}
& (- \nabla^2 + m_s^2) \phi_0 = g_s \rho_s -\frac{1}{2}\kappa \phi_0^2 
-\frac{1}{6} \lambda\phi_0^3; \\
&(-\nabla^2 + m_v^2) V_0 = g_v \rho  -
\frac{1}{3!}\zeta g_v^4 V_0^3 - 2\Lambda \, g_\rho^2 b_0^2 \, g_v^2 V_0; \\
&(-\nabla^2  + m_\rho^2) b_0 = \frac{g_\rho}{2} \rho_3
- 2\Lambda \, g_\rho^2 b_0 \, g_v^2 V_0^2; \\
&-\nabla^2 A_0 = e \rho_p - e \rho_e.
\end{eqnarray}

One can use the above field equations and the fact that the derivatives of the fields are zero at the board of the cells to simplify the expression of the energy. One has:
\begin{eqnarray}
E_{TF} =& \int_V d^3 r~ \left\{ \sum_i \ep_i(\bs{r}) + \f{g_s}{2} \phi_0(\bs{r}) \rho_s(\bs{r}) - \frac{1}{12}\kappa \phi_0^3(\bs{r}) \right. \nonumber \\
& - \frac{1}{24} \lambda\phi_0^4(\bs{r}) + \f{g_v}{2} V_0(\bs{r}) \rho_B(\bs{r}) + \f{\zeta}{24}g_v^4 V_0^4(\bs{r}) \nonumber \\
& + \f{g_\rho}{4} b_0(\bs{r}) \rho_3(\bs{r}) + \Lambda g_\rho^2 \, b_0^2 \, g_v^2 V_0^2  \nonumber \\
&\left. + \f{e}{2} A_0(\bs{r})  [\rho_p(\bs{r})-\rho_e(\bs{r})] \right\}.
\label{eq energia 3}
\end{eqnarray}

Finally, we minimized the thermodynamic grand potential
\begin{equation}\label{eq gran potential}
\Omega=E_{TF}-\sum_{i=p,n,e} \mu_i\int_V d^3r \rho_i({\bs{r}}),
\end{equation}
imposing fixed particle number, where $\mu_i$ are the chemical potentials.
The minimization results:
\begin{eqnarray}
&\mu_p = \sqrt{k_{Fp}^2 + M^{*2}} + g_v V_0(\bs{r}) + \f{1}{2}g_{\rho} b_0(\bs{r}) + e A_0(\bs{r});
\label{eq chem p} \\
&\mu_n = \sqrt{k_{Fn}^2 + M^{*2}} + g_v V_0(\bs{r}) - \f{1}{2}g_{\rho} b_0(\bs{r});
\label{eq chem n} \\
&\mu_e = \sqrt{k_{Fe}^2 + m_e^2} - e A_0(\bs{r}).
\end{eqnarray}

The details of the methods employed to find the solutions as just outlined can be found in \cite{Avancini2008}.

\subsection{Pairing energy}\label{ssec pairing}

The pairing energy, in the BCS approximation, is given by \cite{Tian2009,Ballentine1998}:
\begin{equation}
E_{\textrm{pair}}= -\frac{1}{2} \sum_i \Delta_i u_i v_i
= -\sum_{i>0} \Delta_i u_i v_i,
\end{equation}
where
\begin{eqnarray}
&u_i^2=\frac{1}{2}\left[ 1+\frac{\vep_i-\mu}{[(\vep_i-\mu)^2+\Delta_i^2]^{1/2}} \right]; \\
&v_i^2=\frac{1}{2}\left[ 1-\frac{\vep_i-\mu}{[(\vep_i-\mu)^2+\Delta_i^2]^{1/2}} \right],
\end{eqnarray}
\noindent with $\vep_i$ being the single-particle energies, $\Delta_i$ the pairing gap and $u_i$ and $v_i$ the parameters from the Bogolyubov transformation. One can show that:

\begin{equation}
E_{\textrm{pair}}= -\frac{1}{2}  \sum_{i>0} \frac{\Delta_i^2} 
{[(\vep_i-\mu)^2+\Delta_i^2]^{1/2}} .
\end{equation}
And going to the continuous case:
\begin{equation}\label{eq pair}
E_{\textrm{pair}}= -\frac{1}{4}\frac{\gamma}{(2\pi)^3} \int \frac{\Delta^2 (k)}
{[(\vep(k)-\mu)^2+\Delta^2(k)]^{1/2}} d^3k.
\end{equation}
The gap equation for an interaction $V$ is:
\begin{equation}\label{eq gap bcs}
\Delta_i= -\frac{1}{2} \sum_{j>0} \frac{\braket{i,-i|V|j,-j} \Delta_j}
{[(\vep_j-\mu)^2+\Delta_j^2]^{1/2}}.
\end{equation}

For a particle-particle separable non-local interaction and taking only the S channel, which should be a good approximation
for low energy, one may write: 
\begin{eqnarray}
\braket{i,-i|V|j,-j} &= \braket{k_i s_i,k_{-i} s_{-i}|V^{^1S_0}\frac{1}
{2}(1-P^\sigma)|k_j s_j,k_{-j} s_{-j}}
\nonumber \\
&= \braket{k_i,k_{-i} |\frac{V^{^1S_0}}{2}| k_j, k_{-j}} (\delta_{s_i s_j} \delta_{s_{-i} s_{-j}} - \delta_{s_i s_{-j}} \delta_{s_{-i}s_j}),
\end{eqnarray}
with the $P^\sigma$ operator projecting on to the S channel. We then obtain:
\begin{equation}\label{eq potential}
\braket{i,-i|V|j,-j}= -G p(k_i) p(k_j),
\end{equation}
where $G$ is a constant.

Using the above result in (\ref{eq gap bcs}) and taking its continuous version results:
\begin{equation}
\Delta(k)= -\frac{1}{2}\int_0^\infty \frac{{k'}^2 dk'}{2\pi^2} \frac{-G p(k) p({k'})
	\Delta (k')}
{[(\vep(k')-\mu)^2+\Delta^2(k')]^{1/2}}.
\end{equation}
It is now easy to conclude that $\Delta(k)= \Delta_0 p(k)$ is a solution for the gap equation with $\Delta_0$ being determined by solving the equation:
\begin{equation}\label{eq one}
1= \frac{G}{4\pi^2}\int_0^\infty \frac{p^2(k)}
{[(\vep(k)-\mu)^2+\Delta_0^2 p^2(k)]^{1/2}} k^2 dk.
\end{equation}

In order to obtain the function $p(k)$, we followed the same procedure used in \cite{Tian2009}, i.e.:
\begin{equation}\label{eq pk}
p(k)= \frac{1}{3} \sum_{i=1}^3 e^{-a_i^2 k^2}.
\end{equation}
The parameters $a_i$ and $G$ were then fixed in order to reproduce $\Delta(k_F)$ for the Gogny force \cite{Decharge1980,Tian2009}, solving (\ref{eq one}) for symmetric nuclear matter. In this case, the single-particle energy and chemical potential are $\vep(k) = \sqrt{k^2 + M^{*2}} + g_v V_0$ and $\mu = \sqrt{k_F^2 + M^{*2}} + g_v V_0$, respectively, which are obtained from the Lagrangian (\ref{eq lag}) for nuclear matter. For the pasta phase,
\begin{equation}\label{ener part pasta}
\vep(k) = \sqrt{k^2 + M^{*2}} + g_v V_0 + \frac{\tau_3}{2}g_{\rho} b_0 + \frac{1+\tau_3}{2}e A_0,
\end{equation}
and $\mu$ is given by (\ref{eq chem p}) and (\ref{eq chem n}).
The kinetic particle energy densities and baryonic and scalar densities are now redefined, including the occupation probabilities $v(k)$: 
\begin{eqnarray}
&\ep_i = \frac{1}{\pi^2} \int_0^\infty E_i^* v^2(k) k^2 dk;
\label{eq ener cinetica} \\
&\rho_i = \frac{1}{\pi^2} \int_0^\infty v^2(k) k^2 dk;
\label{eq numero massa ocupacao} \\
&\rho_{si} = \frac{1}{\pi^2} \int_0^\infty \f{M^*}{E_i^*} v^2(k) k^2 dk,
\label{eq rhos ocupacao}
\end{eqnarray}
with
\begin{equation}\label{ocupacao}
v^2(k)=\frac{1}{2}\left[ 1-\frac{\vep(k)-\mu}{[(\vep(k)-\mu)^2+\Delta^2(k)]^{1/2}} \right].
\end{equation}

In a finite system, the fields and the above quantities ((\ref{ener part pasta}) to (\ref{ocupacao})) are position dependent, therefore the gap is also position dependent via (\ref{eq one}).
In this case (\ref{eq pair}) have an extra $\bs{r}$ integration.
Finally, the total energy is given by $E_{TF}+E_{\textrm{pair}}$.

Although the fit of the parameters $a_i$ and $G$ was made only for symmetric matter and there is no isospin dependence in $G$ and $p(k)$, the pairing gap trend for neutron-rich matter is taken into account via the isospin dependence of $\vep(k)$, which reflects the final value and behaviour of $\Delta_0(r)$, obtained self-consistently in the TF procedure.

\section{Results}\label{resultados}

Based on the above formal results, we may include the pairing in the pasta phase calculation in two different ways. In the first one, we simply solve the pasta and take the solutions of the field equations to use those fields to solve the gap equation and then calculate the occupation probabilities. These probabilities are then used to redefine the densities and to obtain the total energy $E_{TF}+E_{\textrm{pair}}$. We call this the approach NSC.
As a second approach, we solve the field equations including the pairing effect since the beginning of the variational procedure. In this case (which we call SC) the field and gap equations are solved simultaneously. We concentrated our results on the SC approach and presented a comparison with NSC in the end.

 \begin{figure}
 	\begin{center}
 		\includegraphics[scale=1.]{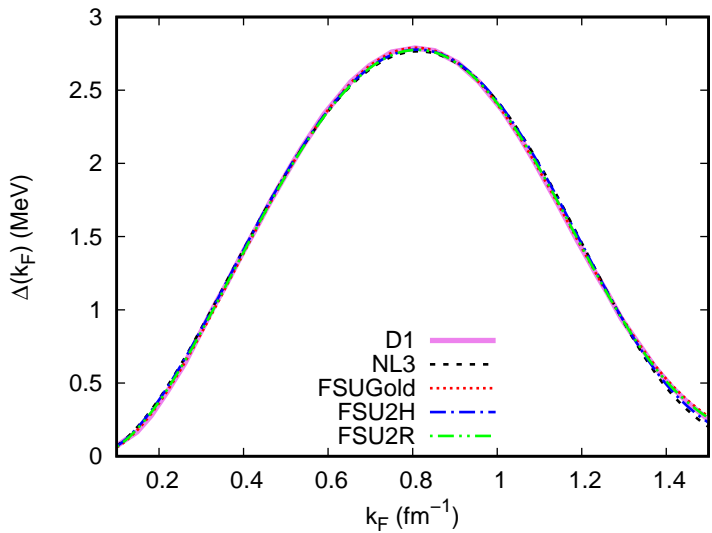}
 	\end{center}
 	\caption{Gap $\Delta(k_F)$ for symmetric nuclear matter for the parametrizations of table \ref{tab2 models}, adjusted to fit the Gogny D1 interaction (see details in \cite{Tian2009}). The parameters obtained are given in table \ref{tab1 delta}.}
 	\label{fig1 delta}
 \end{figure}

\begin{table}
  \centering
        \caption{Parameters from (\ref{eq potential},\ref{eq pk}).}
	\begin{tabular}{lcccc}
                \hline
                \hline
		 & $G$ (MeV fm$^{-3}$) & $a_1$ (fm) & $a_2$ (fm) & $a_3$ (fm) \\
                \hline
		NL3     & 600.40 & 0.629 & 1.293 & 0.152 \\
		FSUGold & 728.01 & 0.454 & 1.126 & 0.454 \\
		FSU2H   & 583.17 & 0.633 & 1.252 & 0.145 \\
		FSU2R   & 679.11 & 0.614 & 1.198 & 0.233 \\
		\hline
		\hline
	\end{tabular}
	\label{tab1 delta}
\end{table}

\begin{table}
  \centering
        \caption{Parametrization properties for infinite symmetric nuclear matter
        at zero temperature and
saturation density $\rho_0$: binding energy per nucleon $E_B/A$, incompressibility $K$, nucleon effective mass $M^\ast$, symmetry energy at saturation density $E_{sym}$ and its slope $L$.}
	\begin{tabular}{lccccccc}
                \hline
                \hline
		 & $\rho_0$ & $E_B/A$ & $K$ & $M^\ast/M$ & $E_{sym}$ & $L$ \\
		 & (fm$^{-3})$ & (MeV) & (MeV) & & (MeV) & (MeV) \\
                \hline
		NL3 & 0.148 & -16.299 & 271.76 & 0.6 & 37.4 & 118.3 \\
		FSUGold & 0.148 & -16.30 & 230.0 & 0.620 & 32.6 & 60.5 \\
		FSU2H & 0.1505 & -16.28 & 238.0 & 0.593 & 30.5 & 44.5 \\
		FSU2R & 0.1505 & -16.28 & 238.0 & 0.593 & 30.7 & 46.9 \\
		\hline
		\hline
	\end{tabular}
	\label{tab2 models}
\end{table}

 In figure \ref{fig1 delta} and table \ref{tab1 delta} we present our results for the gap $\Delta(k_F)$ and the numerical values for the parameters $G$ and $a_i$, respectively, obtained by fitting the Gogny interaction
 for symmetric nuclear matter for the parametrizations mentioned in section \ref{formalismo}, which properties are presented in table \ref{tab2 models}. Note that, in the NL3 case, our parameters are different from \cite{Tian2009}, although our fit is very good in this case too.

\begin{figure}
\begin{center}
\includegraphics[scale=1.]{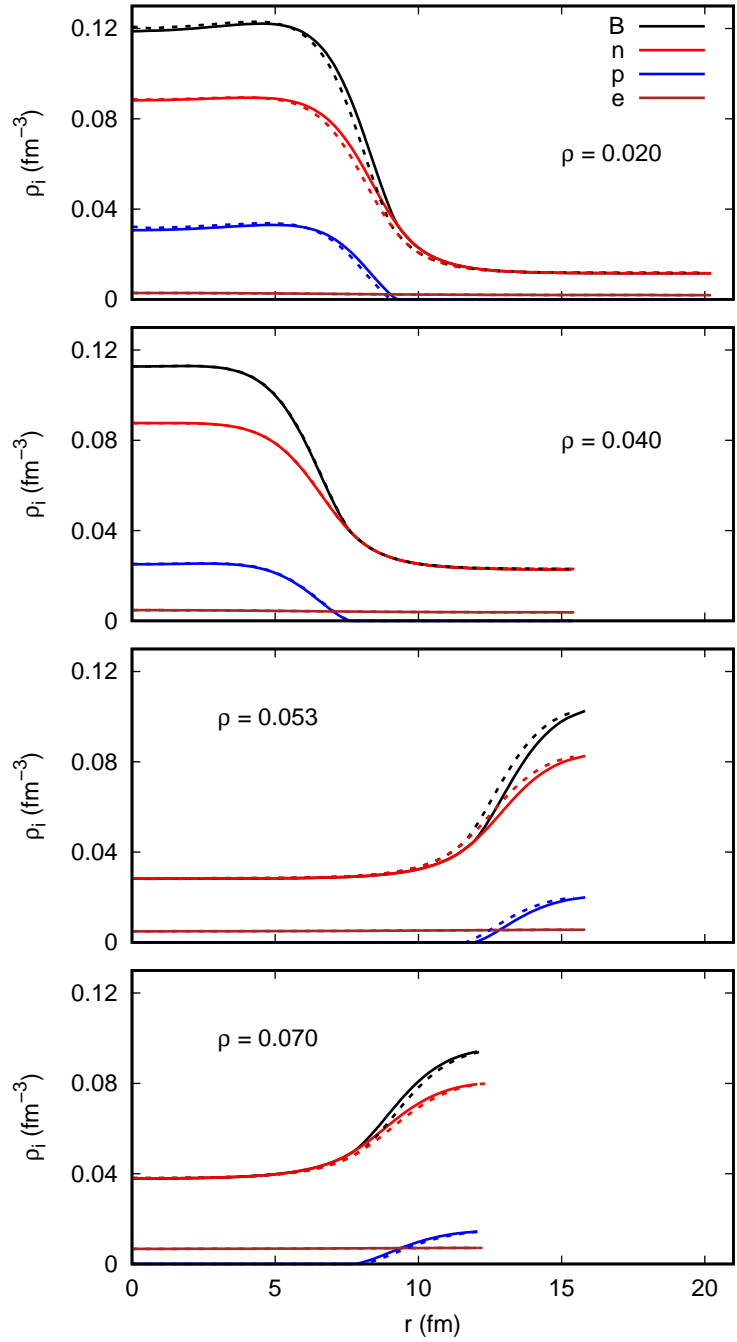}
\end{center}
\caption{Density profiles (baryon, proton, neutron and electron) for FSUGold, $y_p=0.1$ and different densities (in fm$^{-3}$, indicated in the figure). Solid lines without pairing; dashed lines for SC. From top to bottom panels, one has droplet, rod, tube and bubble.}
\label{fig2 RHOs}
\end{figure}
 
\begin{figure}
\begin{center}
\includegraphics[scale=1.]{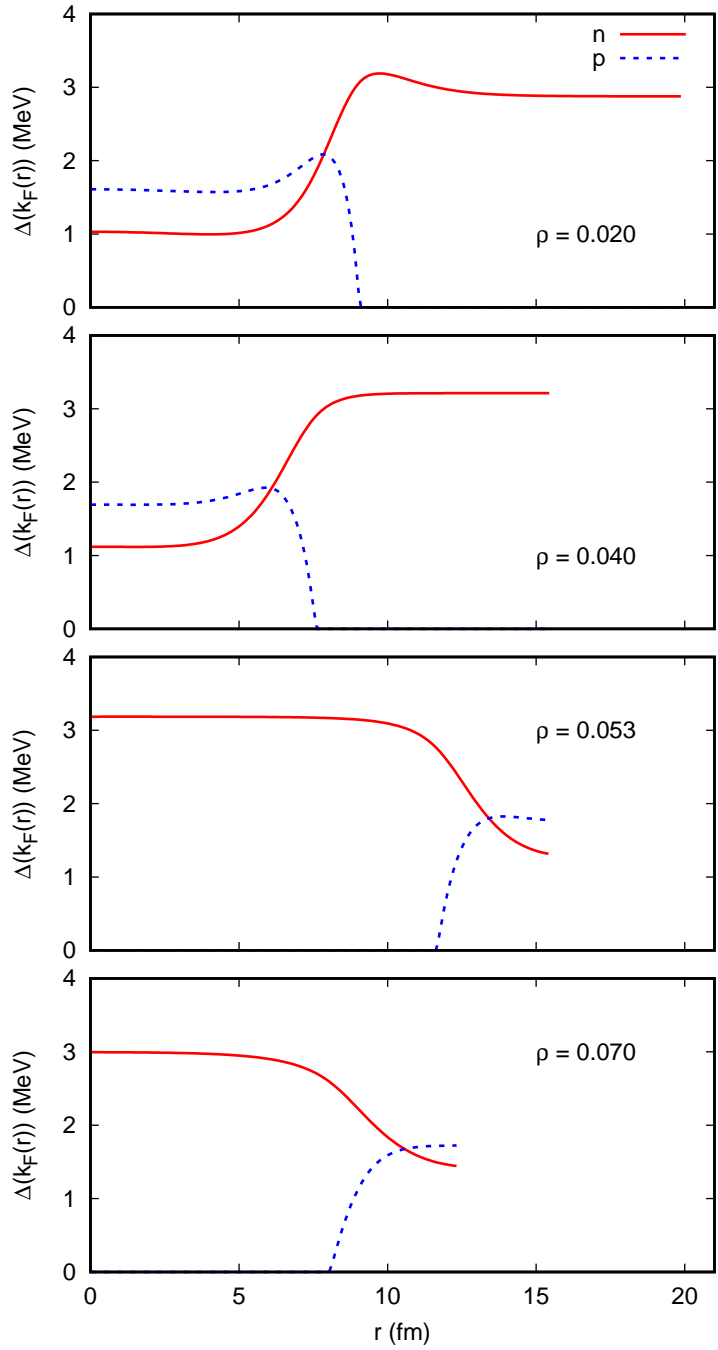}
\end{center}
\caption{$\Delta(k_F(r))$ for protons and neutrons for FSUGold, $y_p=0.1$ and different densities (in fm$^{-3}$, indicated in the figure).
From top to bottom panels, one has droplet, rod, tube and bubble.}
\label{fig3 DELTAs}
\end{figure}

\begin{figure}
\begin{center}
\includegraphics[scale=1.]{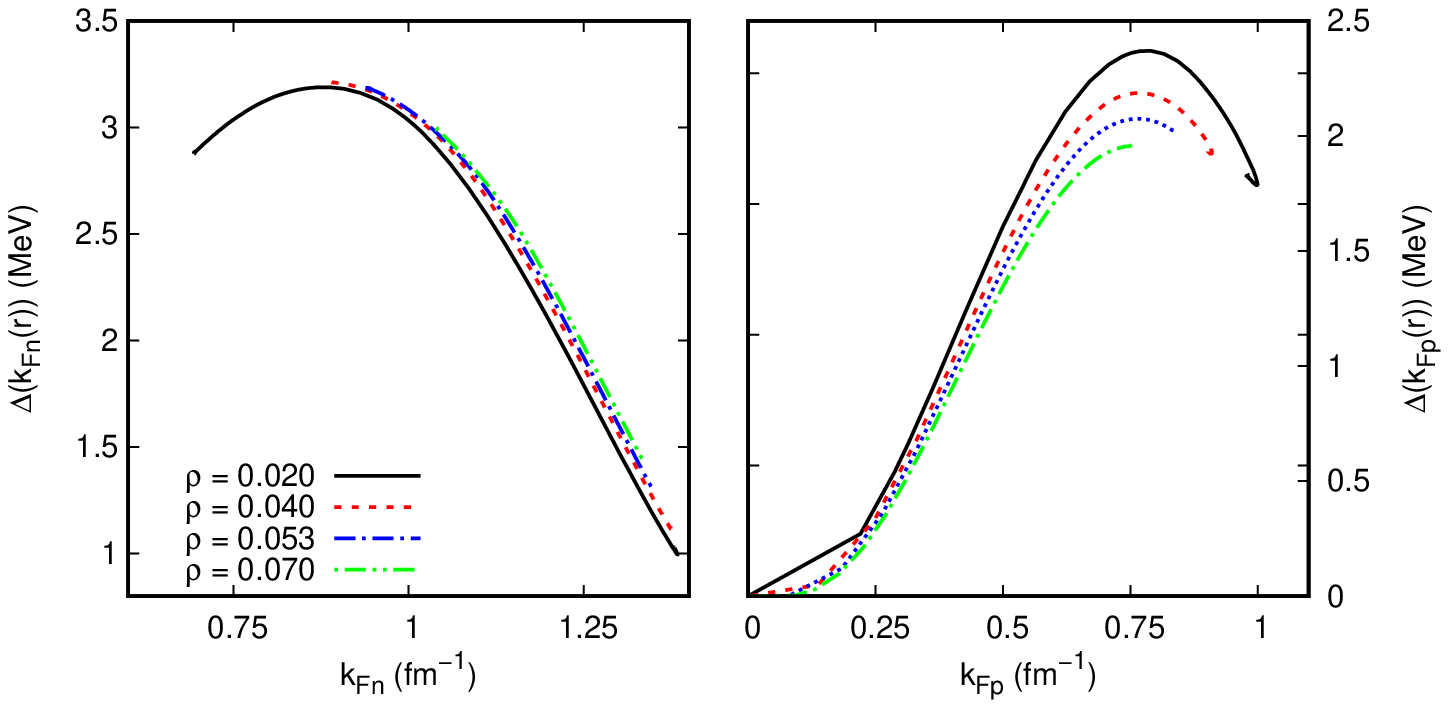}
\end{center}
\caption{Same as figure \ref{fig3 DELTAs}, but as a function of the Fermi momenta.}
\label{fig4 DELTAs_vkf}
\end{figure}

Once in possession of those parameters, we solved the gap equation (\ref{eq one}) for the finite TF case. In figure \ref{fig2 RHOs} we present the results for the particle density profiles throughout the WS cells for some global densities. At these global densities one has different pasta structures as the most stable, although this depends on the parametrization and on the proton fraction. In particular, for $y_p=0.1$
and FSUGold, the slab structure does not appear. Since the pairing is a small effect, its inclusion almost does not affect the density profiles, including the WS radii. Our results for $\Delta(k_F(r))$ as a function of the position and of $k_F(r)$ are shown in figures \ref{fig3 DELTAs} and \ref{fig4 DELTAs_vkf} for the same parametrization and proton fraction as in figure \ref{fig2 RHOs}.
As expected, for intermediate proton or neutron densities one has correspondingly larger $\Delta(k_F(r))$ (see figure \ref{fig1 delta}). Because of the isospin dependence of $\vep(k)$ in (\ref{eq one}), which is not the same as for symmetric matter, values of $\Delta(k_F(r))>3$ MeV for the neutrons are found. This behaviour is in accordance with the one presented in \cite{Pastore2011} for the Gogny force.
The behaviour is similar for the other parametrizations.

\begin{figure}
 	\begin{center}
 		\includegraphics[scale=1.]{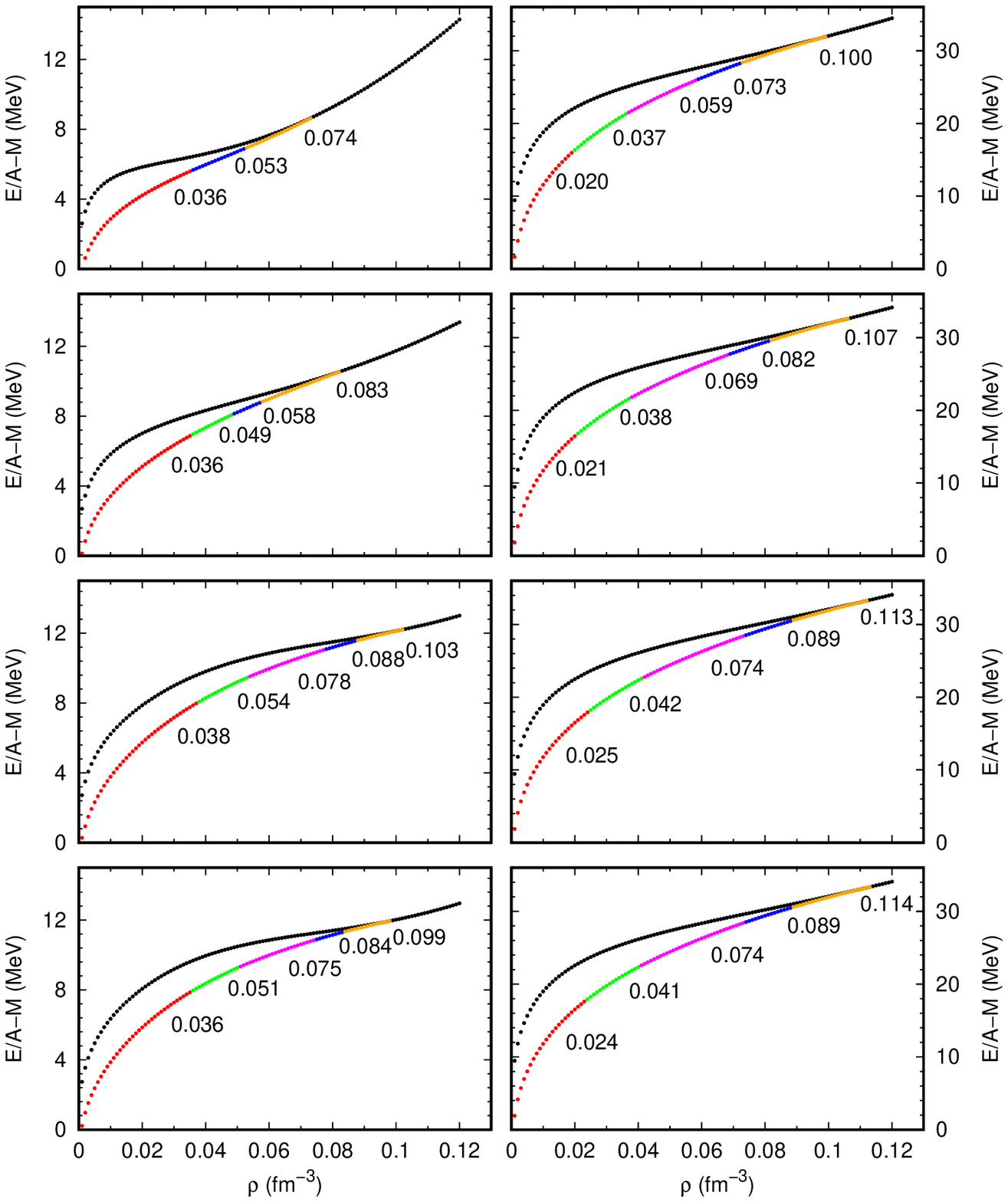}
 	\end{center}
 	\caption{Energy per nucleon vs total density for $y_p=0.1$ (left side) and $y_p=0.3$ (right side). From top to bottom panels, one has
 	NL3, FSUGold, FSU2H and FSU2R. Only the energy for the most stable pasta structure is shown. From left to right, one has droplet (red), rod (green), slab (pink), tube (blue) and bubble (orange). Black is for homogeneous matter. Note that sometimes not all structures appear. The transition densities are indicated in the figures.}
 	\label{fig5 Ener_v2}
 \end{figure}

In figure \ref{fig5 Ener_v2}
one can see the comparison between the energy per nucleon in the WS cell and in the homogeneous matter.
One can see that one has a dependence with the parametrization and with the proton fraction.
The uniform matter result, as expected, has a larger free energy than the pasta phases. Although the transition densities between geometries differ, depending on the
parametrization used, the qualitative behaviour does not change significantly for $y_p=0.3$. For $y_p=0.1$, the parametrization dependence is more visible, being FSU2H and FSU2R close to each other, which could be expected since these two parametrizations are similar, as can be seen from the main properties depicted in table
\ref{tab2 models}.

\begin{figure}
 	\begin{center}
 		\includegraphics[scale=1.]{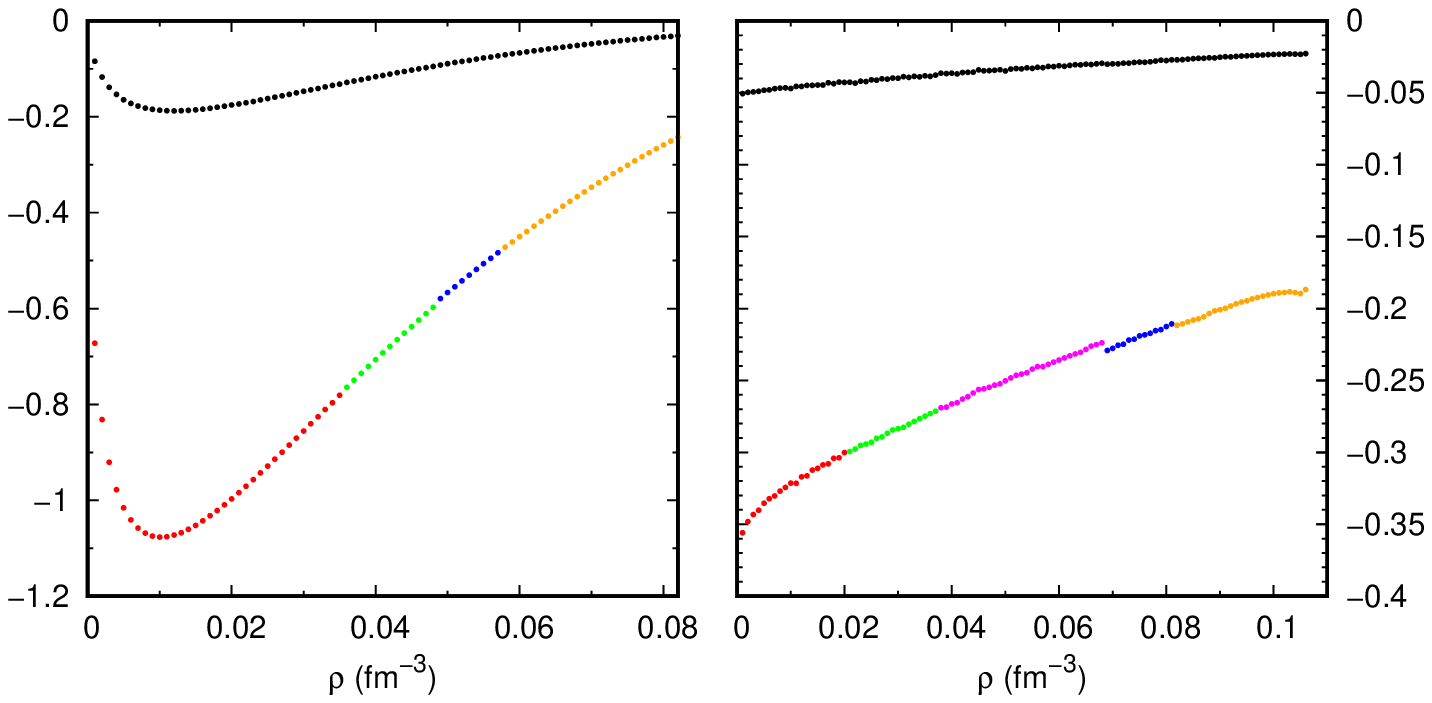}
 	\end{center}
 	\caption{Total energy per nucleon difference between the cases with and without pairing, $(E/A)-(E/A)_\textrm{wp}$ (top black dots, in MeV), and pairing energy per nucleon, $E_{\textrm{pair}}/A$ (bottom colorful dots, also in MeV). From left to right, one has droplet (red), rod (green), slab (pink), tube (blue) and bubble (orange). Left panel is for $y_p=0.1$ and right panel is for $y_p=0.3$. FSUGold parametrization was used.}
 	\label{fig6 Epair}
 \end{figure}

Figure \ref{fig6 Epair} shows the pairing energy per nucleon, $E_{\textrm{pair}}/A$, and the total energy per nucleon difference between the cases with and without pairing, $(E/A)-(E/A)_\textrm{wp}$, both for FSUGold. Because of the occupation probabilities in (\ref{eq ener cinetica}), (\ref{eq numero massa ocupacao}) and (\ref{eq rhos ocupacao}), $E_{\textrm{pair}}/A$ is not the difference in the total energy per particle between the cases with and without pairing. One can see that the difference between the total energies increases with the global asymmetry.

Some previous works have studied the influence of the symmetry energy $E_\textrm{sym}$ and its slope $L$ on the pasta phase, as, for instance, in \cite{Xia2021,Bao2014}. They showed that the transition densities between the different pasta structures and also the transition pasta-homogeneous matter is strongly affected by $L$. In particular, a smaller $L$ increases the transition pasta-homogeneous matter for asymmetric matter. We calculated the symmetry energy with the pairing in all the density range of interest here  and we found a difference less than 0.1 MeV in the interval. As a consequence, we found a very small difference for $L$ at saturation density. As an example, for FSUGold $L=60.33~(60.38)$ MeV with (without) pairing. Since $E_\textrm{sym}$ and $L$ changes very little with the inclusion of pairing, we expect that transition densities also change very little between the different pasta structures. For example, for FSUGold and $y_p=0.1$, with pairing (without pairing) one has $\rho=0.036~(0.036)$ fm$^{-3}$ for droplet-rod, $\rho=0.049~(0.049)$ fm$^{-3}$ for rod-tube, $\rho=0.058~(0.057)$ fm$^{-3}$ for tube-bubble and $\rho=0.083~(0.083)$ fm$^{-3}$ for bubble-homogeneous matter.

\begin{figure}
	\begin{center}
	\includegraphics[scale=1.]{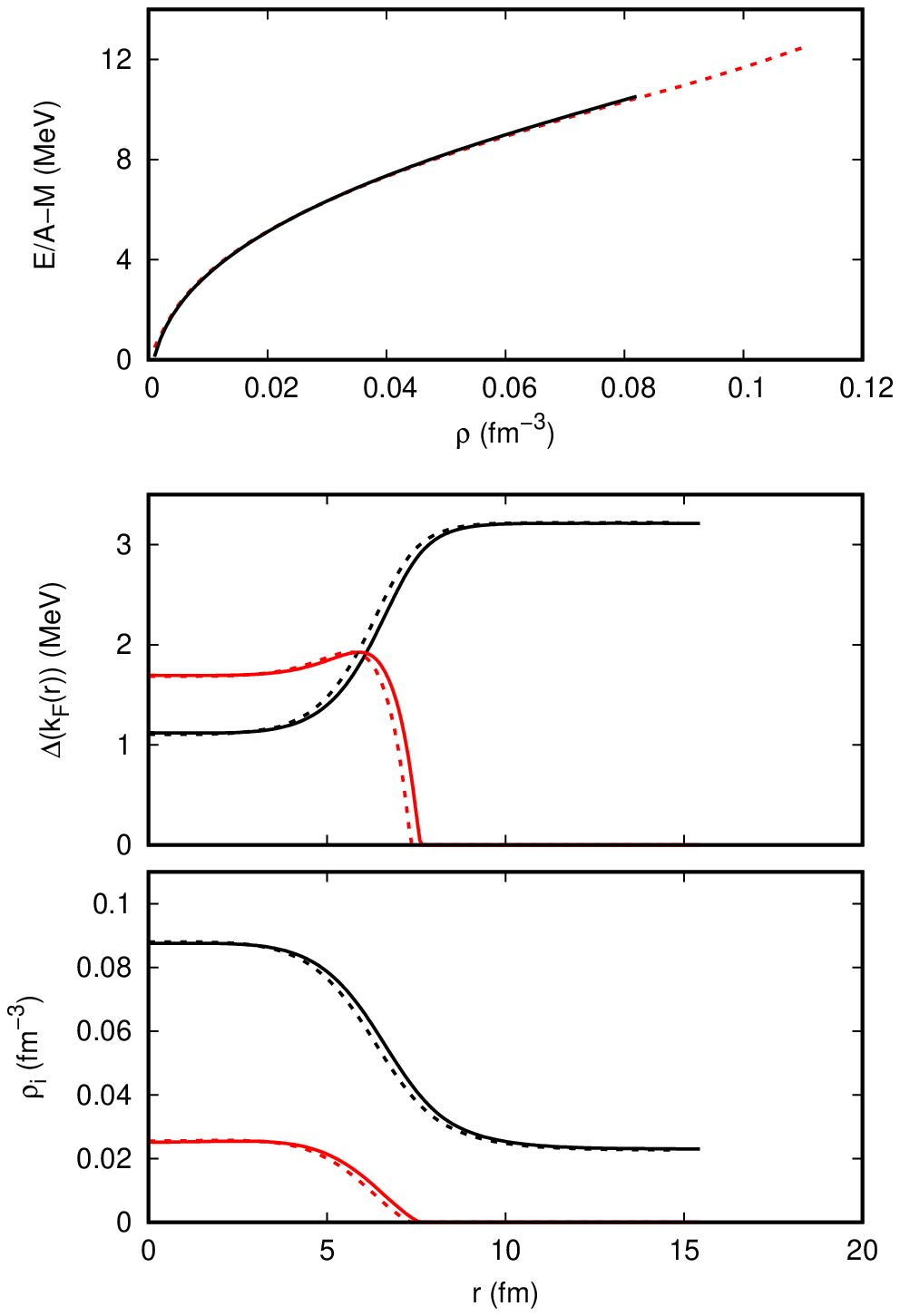}
	\end{center}
	\caption{From top to bottom: energy per nucleon, neutron (black line)
and proton (red line) gaps $\Delta(k_F(r))$ and neutron and proton density profiles. The last two quantities are shown for global density $\rho=0.04\mathrm{~fm}^{-3}$. The solid line is the result for the approach SC and the dashed one for NSC. The chosen parametrization is FSUGold with $y_p=0.1$.}
	\label{fig7 comp_SCxNSC}
\end{figure}

In figure \ref{fig7 comp_SCxNSC} we compare the two approaches mentioned in the beginning of the section, SC and NSC, for a particular case. As we can see, the resulting equation of state is essentially the same, while small differences can be noticed for the quantities $\Delta(k_F(r))$ and  $\rho(r)$. These small differences are also present for different parametrizations and total densities. Still concerning the differences between SC and NSC, in table \ref{tab3 transition} we show the densities where there is the transition from the pasta to homogeneous matter, including the case where there is no pairing at all, for the parametrizations considered in this work.

\begin{table}
	\centering
	\caption{Transition densities (in fm$^{-3}$) between pasta phase and homogeneous matter in the approaches NSC and SC and for the no pairing case, for each model parameter set used and two different proton fractions.}
	\begin{tabular}{lcccc}
		\hline
		\hline
		& ~~$y_p$~~ & ~~NSC~~ & ~~SC~~ & NO PAIR  \\
		\hline
		NL3 & 0.1 & $0.074$ & $0.074$ & $0.074$ \\
		FSUGold & 0.1 & $0.079$ & $0.083$ & $0.083$ \\
		FSU2H & 0.1 & $0.100$ & $0.103$ & $0.102$ \\
		FSU2R & 0.1 & $0.097$ & $0.099$ & $0.099$ \\
		NL3 & 0.3 & $0.099$ & $0.100$ & $0.100$ \\
		FSUGold & 0.3 & $0.101$ & $0.107$ & $0.104$ \\
		FSU2H & 0.3 & $0.112$ & $0.113$ & $0.113$ \\
		FSU2R & 0.3 & $0.111$ & $0.114$ & $0.112$ \\
		\hline
		\hline
	\end{tabular}
	\label{tab3 transition}
\end{table}

As is well known, the inner crust of neutron stars, contains  a non-uniform system for which the model described above should be suited. In this part of the star there is a strong temperature drop due to neutrino emission in the core to the surface. The rate of the conveyance of energy depends on the specific heat and on the thickness of the inner crust. Since the specific heat is sensitive to superfluidity, we may now, with the inclusion of pairing as just implemented, obtain that quantity straightforwardly. We follow here the approach used in \cite{Barranco1998} and \cite{Nakano2001}, which starts with the definition of the specific heat:

\begin{equation}
C_{v}=\frac{1}{V} \frac{\partial <E>}{\partial T},
\end{equation}

\noindent with $V$ being the WS cell volume and $T$ the temperature. For the mean quasiparticle energy $<E>$, we use the definition:

\begin{equation}
<E>=\int \frac{d^{3}kd^{3}r}{2\pi^{3}}n_{k}(r)E_{k}(r),
\end{equation}

\noindent where $n_{k}=1+\exp(E_{k}(r)/T)$ and $E_{k}(r)=[(\sqrt{k^2+M^*(r)^2}-\sqrt{k_F^2+M^*(r)^2})^{2}+\Delta^{2}_{k}(r)]^{1/2}$. Disregarding now the term $\frac{\partial E_{k}}{\partial T}$, which should be a good approximation for small temperatures \cite{Barranco1998}, we get the expression:

\begin{equation}
C_{v}=\frac{1}{V}.\frac{\gamma}{8\pi^{2}T^{2}}.\int d^3r \int k^{2}dk \frac{E_{k}^{2}}{cosh^{2}(E_{k}/2T)},
\end{equation}

\noindent where the differential $d^3r$ will depend on the particular geometry of the pasta phase.

\begin{figure}
	\begin{center}
		\includegraphics[scale=1.0]{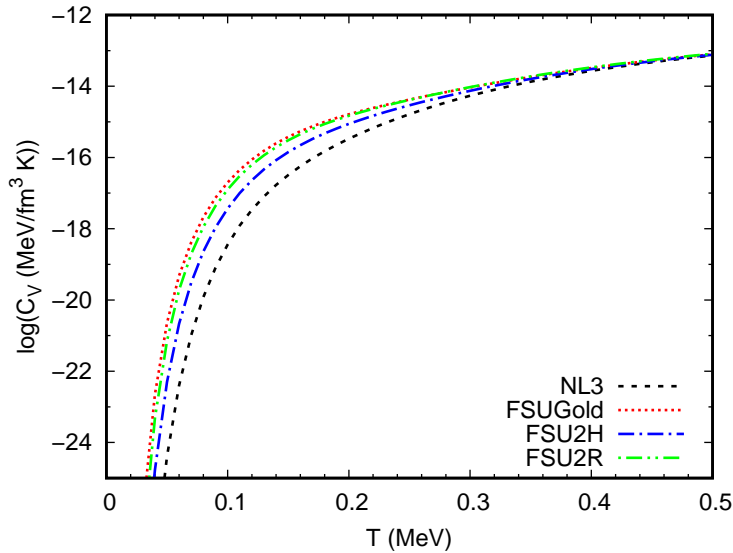}
	\end{center}
	\caption{Specific heat of the neutrons as a function of the temperature for the various parametrizations used in this paper, for $y_p=0.1$ and a global density $0.017\mathrm{~fm}^{-3}$, corresponding to a droplet geometry.
	}
	\label{fig8 cal_parameters}
\end{figure}

\begin{figure}
	\begin{center}
		\includegraphics[scale=1.0]{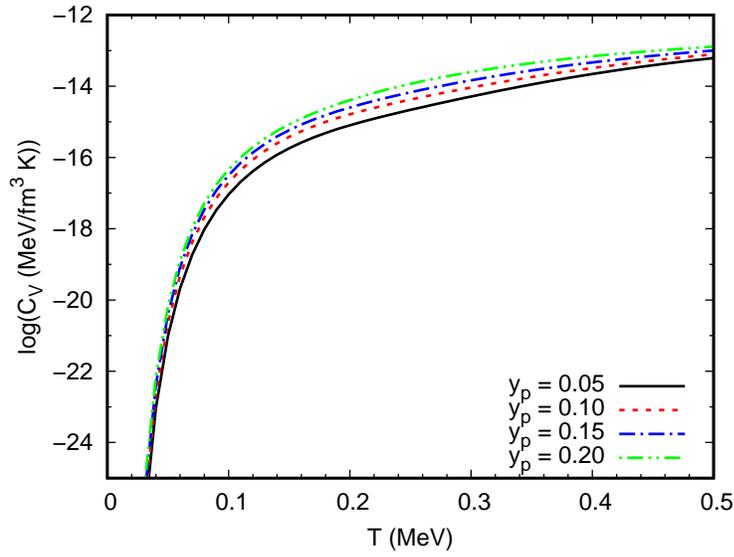}
	\end{center}
	\caption{Specific heat of the neutrons for the FSUGold case and $\rho=0.017\mathrm{~fm}^{-3}$, for four different total proton fractions.}
	\label{fig9 cal_fraction}
\end{figure}

\begin{figure}
	\begin{center}
		\includegraphics[scale=1.0]{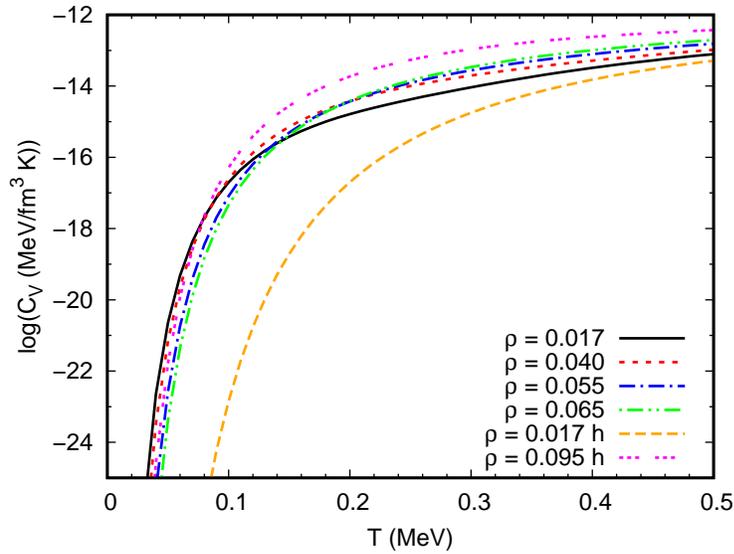}
	\end{center}
	\caption{Specific heat of the neutrons for different geometries and FSUGold with $y_p=0.1$: $\rho=0.017\mathrm{~fm}^{-3}$ droplet, $\rho=0.040\mathrm{~fm}^{-3}$ rod, $\rho=0.055\mathrm{~fm}^{-3}$ tube, $\rho=0.065\mathrm{~fm}^{-3}$ bubble and the last two $\rho=0.017\mathrm{~fm}^{-3}$ and $\rho=0.095\mathrm{~fm}^{-3}$ are for homogeneous matter (indicated with ``h'').}
	\label{fig10 cal_structure}
\end{figure}

\begin{figure}
	\begin{center}
		\includegraphics[scale=1.0]{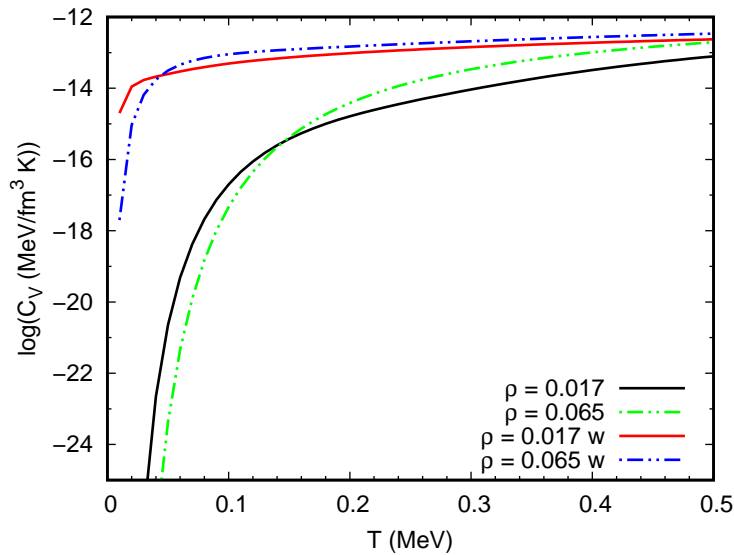}
	\end{center}
	\caption{Specific heat of the neutrons for FSUGold with $y_p=0.1$. $\rho=0.017\mathrm{~fm}^{-3}$ droplet, $\rho=0.065\mathrm{~fm}^{-3}$ bubble. ``w'' stands for pasta without pairing.}
	\label{fig11 cal_wp}
\end{figure}

In figures 
\ref{fig8 cal_parameters}, \ref{fig9 cal_fraction} and \ref{fig10 cal_structure} we display our results for the specific heat of the neutrons
comparing the different parametrizations used here at the same global density and proton fraction, the results for different proton fractions with the same parametrization and density, and for different geometries and same parametrization and proton fraction, respectively. In figure \ref{fig10 cal_structure} we also include our results for homogeneous matter.
In figure \ref{fig11 cal_wp} we compare the specific heat of the neutrons with and without pairing for the pasta phase.

From figure \ref{fig8 cal_parameters} one can see a strong dependence for very low temperatures with the model parametrization, since the logarithm of the specific heat is displayed. Even for FSU2R and FSU2H, which provide almost equal properties in table \ref{tab2 models}, the coupling constants are different enough to produce the
difference observed in the specific heat. As the temperature increases, the model parametrization dependence decreases.

There is also a dependence with the total proton fraction (figure \ref{fig9 cal_fraction}):
$C_V$ increases with $y_p$. As for different densities the pasta structures are different, the dependence is neither increasing nor decreasing (figure \ref{fig10 cal_structure}). In comparison with the specific heat for homogeneous matter, one can see a big difference for small temperatures.

The inclusion of the pairing strongly affects the specific heat for small temperatures. From figure \ref{fig11 cal_wp} it can be seen that the difference extends to temperatures higher than those considered here and therefore calculations of the specific heat of the pasta including the temperature self-consistently need to be done. The inclusion of the pairing in the calculation of the specific heat significantly influences the thermalization of neutron star crusts \cite{Fortin2010}, enhancing the cooling at the surface of the star.

\section{Conclusions}\label{finale}

In the present work we implemented the pairing contribution to a Thomas-Fermi calculation, previously investigated and applied to the non-uniform nuclear pasta phase. To this end we first found a series of expressions to describe the well known Gogny gap for the pairing interaction in the nuclear homogeneous matter. Those expressions were obtained by a fitting procedure for four different parametrizations of the model Lagrangian and they show an almost perfect accordance with Gogny's interaction in the density region of interest. We then used those fitted expressions to obtain the position dependence of the gap, needed for a non-uniform calculation. This last procedure was done in two different ways: solving the gap equation after the minimization of the grand potential, as explained before (\ref{eq gran potential}) in the text, and as a second approach, solving the gap equation self-consistently with the grand potential minimization. The differences encountered for the gap and the final density profiles are small though noticeable.

Our results indicate a very small modification in the equation of state for the pasta phase with the inclusion of the pairing, although a slight change in some transition points between the various phases in the pasta could be noticed, specially the transition between non-uniform and uniform matter phases when the NSC approach is used. Noticeable differences could also be founded at this respect when we compared the two approaches mentioned here for the introduction of the pairing in our TF calculation.

An important quantity which is very sensitive to pairing effects is the specific heat. Albeit our calculation in this paper does not include explicitly temperature effects, it is possible to obtain the behavior of that quantity for small temperatures. We presented our results for the specific heat of the neutrons in the pasta phase, considering different parametrizations for the model Lagrangian, different pasta geometries and different proton fractions for values that are expected in the inner crust of neutron stars. Our results are consistent with other previous similar calculations. For a quantitative comparison with more recent calculations \cite{Pastore2014a,Pastore2014b}, the explicit inclusion of temperature effects should be done. Our previous work on that matter \cite{Avancini:2010ch} gives us some confidence that those effects, including now pairing, is straightforward. In particular, comparison with results using full HFB calculations should be of interest, once the TF demands much less calculational cost, even for very large number of particles.

As discussed in previous works, the pasta phase is important to supernovae and evolution of neutron stars. Although the pairing produces little modification in the equation of state, it influences other quantities relevant to such objects. The specific heat with pairing is expected to enhance the thermalization of neutron star crusts \cite{Fortin2010}.
Some works have argued that neutrino transport and emissivity are also affected by pairing \cite{Burrello2016,Sedrakian:2018ydt}. The thermalization could therefore be affected again, as well as the chemical composition during various stages of CCSN and neutron stars. Neutrino cross-sections within the formalism presented here will be investigated in a future work.

\ack

We acknowledge partial support from CAPES and CNPq.

\section*{References}
\bibliography{bibli}
\bibliographystyle{unsrt}

\end{document}